\documentclass[fleqn,usenatbib]{mnras}

\usepackage{newtxtext,newtxmath}

\usepackage{ulem}
\usepackage{nicefrac}

\usepackage[T1]{fontenc}
\usepackage{ae,aecompl}

\usepackage{graphicx}	
\usepackage{amsmath}	
\usepackage{amssymb}	
\graphicspath{{figure/}} 
\usepackage{lipsum}
\usepackage{graphicx}
\usepackage{xcolor}
\usepackage{url}
\usepackage{multicol}
\usepackage{multirow}
\usepackage{mathtools}
\usepackage{arydshln}
\usepackage{enumerate}
\usepackage{hyperref}
\usepackage{blindtext}
\usepackage{bm}
\usepackage{amsbsy}
\usepackage{fixmath}
\usepackage[english]{babel}
\usepackage{float}
\usepackage{caption}
\usepackage{gensymb}
\usepackage[utf8]{inputenc}
\usepackage{fixltx2e}

\def\simgt{\lower.5ex\hbox{$\; \buildrel > \over \sim \;$}}

\def\aap{A\&A}
\def\apj{ApJ}
\def\mnras{MNRAS}
\def\apjs{ApJS}

\def\fdm{f_{\rm DM}}

\title[SEAGLE-III: The central dark matter fractions in simulated and observed galaxies]{SEAGLE--III: Towards resolving the mismatch in the dark-matter fraction in early-type galaxies between simulations and observations} 

\author[Mukherjee et al.]{Sampath Mukherjee$^{1,2}$\thanks{\href{mailto:sampath.mukherjee@uliege.be}{\nolinkurl{sampath.mukherjee@uliege.be}}},
L\'{e}on V. E. Koopmans$^{1}$, Crescenzo Tortora$^{3}$, \newauthor Matthieu Schaller$^{4,5}$,  R. Benton Metcalf$^{6,7}$, Joop Schaye$^{5}$, Georgios Vernardos$^{8}$\\
\\
$^{1}$Kapteyn Astronomical Institute, University of Groningen, PO Box 800, 9700AV Groningen, The Netherlands\\
$^{2}$STAR Institute, Quartier Agora - All\'{e}e du six Ao$\hat{u}$t, 19c B-4000 Li\`ege, Belgium \\
$^{3}$INAF -- Osservatorio Astronomico di Capodimonte, Salita Moiariello 16, 80131 - Napoli, Italy\\
$^{4}$Lorentz Institute for Theoretical Physics, Leiden University, PO Box 9506, NL-2300 RA Leiden, The Netherlands\\ 
$^{5}$Leiden Observatory, Leiden University, PO Box 9513, 2300 RA Leiden, The Netherlands\\
$^{6}$Dipartimento di Fisica e Astronomia, Universit\`a di Bologna, via Gobetti 93/2, I-40129 Bologna, Italy\\
$^{7}$INAF - Osservatorio di Astrofisica e Scienza dello Spazio di Bologna, via Gobetti 93/3, I-40129 Bologna, Italy\\
$^{8}$Institute of Physics, Laboratory of Astrophysique, École Polytechnique Fédérale de Lausanne (EPFL), Observatoire de Sauverny, 1290 Versoix, Switzerland
}
\date{Accepted XXX. Received YYY; in original form ZZZ}
\pubyear{2020}
\begin{document}
\label{firstpage}
\pagerange{\pageref{firstpage}--\pageref{lastpage}}
\maketitle


\begin{abstract}
 \noindent
The central dark-matter fraction of galaxies is sensitive to feedback processes during galaxy-formation. Strong gravitational lensing has been effective in the precise measurement of the dark-matter fraction inside massive early-type galaxies.    
Here, we compare the projected dark-matter fraction of early-type galaxies inferred from the SLACS strong-lens survey, with those obtained from the EAGLE, Illustris, and IllustrisTNG hydro-dynamical simulations. 
Previous comparisons with some simulations revealed a large discrepancy, with considerably higher inferred dark-matter fractions -- by factors $\approx$2-3 --  inside half of the effective radius in observed strong-lens galaxies as compared to simulated galaxies.  Here, we report good agreement between EAGLE and SLACS for the dark-matter fractions inside both half of the effective radius and the effective radius as a function of the galaxy's stellar mass, effective radius, and total mass-density slope. However, for IllustrisTNG and Illustris, the dark-matter fractions are lower than observed. This work consistently assumes a Chabrier IMF, which suggests that a different IMF (although not excluded) is not necessary to resolve this mismatch. The differences in the stellar feedback model between EAGLE and Illustris and IllustrisTNG, are likely the dominant cause of the difference in their dark-matter fraction, and density slope.
\end{abstract}


 \section{Introduction}\label{intro}

The study of massive early-type galaxies (ETGs) has been of central interest in galaxy evolution and cosmological studies. ETGs are believed to be the end product of hierarchical galaxy formation (e.g.\ \citealt{toomre1972,cole2000}). The determination of the distribution of baryonic and dark matter within these galaxies is a key step in addressing open questions in galaxy formation and evolution (e.g. \citealt{cappellari2006,koopmans2009, auger2010b,Napolitano2010,tortora2012, tortora2018,lovell2018}). 
One of the still not well-understood issues originating from the gravitational interplay between baryonic and dark matter during galaxy formation  results in the so-called `bulge-halo conspiracy' referring to the total mass distribution following a nearly isothermal profile while the baryons and dark matter, individually, do not \citep{treu2006, Humphrey10, cappellari2006, Thomas2011, Tortora14}. The evidence for this has been reported in strong and weak lensing \citep{treu2004, gavazzi2007, auger2010b} and in stellar-dynamics studies \citep{Dutton_Treu14,Tortora14,LiR+20_MANGA}. 
However, a particular choice for the stellar initial mass function (IMF) is often assumed because it is usually ill-determined.
Furthermore, two parameters have often been used to quantify the distribution of dark matter: the dark-matter fraction within a given radius  ($f_{\rm DM}$) and the logarithmic slope ($\gamma$) of the total density profile.  
To correctly infer $f_{\rm DM}$ from simulations, one therefore additionally requires the simulations to yield $\gamma \approx 2$ -- the isothermal case -- for the combination of baryonic and dark matter both of which are individually non-isothermal \citep{auger2010b,Tortora14}. 

Until recently, cosmological hydrodynamic simulations have been unable to simultaneously match both the observed $f_{\rm DM}$ and $\gamma$ distributions. The isothermal density slopes can be reproduced in simulations by having no or weak feedback, but this leads to an overestimated galaxy formation efficiency and an underestimated dark-matter fraction \citep{Duffy2010}. Conversely, reproducing the observed dark-matter fraction requires strong feedback, but predicts total density slopes that are more shallow (smaller $\gamma$) than isothermal \citep{Dubois2013}. 

For example, \citet{xu2017} studied elliptical galaxies from the \textsc{Illustris} hydrodynamic simulations \citep{vogel2014}, finding lower dark-matter fractions and steeper density slopes inside half of an effective radius than those observed in elliptical galaxies \citep{auger2009, auger2010a}.  A similar trend is found in \textsc{IllustrisTNG} (see middle panel of figure 9 in \citealt{Wang_2019}.). \cite{remus2017} found that the $f_{\rm DM}$ in the \textsc{Magneticum} simulations is lower than observed\footnote{\cite{remus2017} compared $f_{\rm DM} (<R_{\rm eff})$ from \textsc{Magneticum} to $f_{\rm DM} (<R_{\rm eff}/2)$  in SLACS. Although not correct, comparing at the same radius would not resolve the discrepancy as shown by \cite{Wang_2019}}. 
These mismatches 
point to either an inadequacy in the theoretical model or systematic biases in the observational methods. Having found lower $f_{\rm DM}$ values in IllustrisTNG than in SLACS, \cite{Wang_2019}  
suggest that a Salpeter IMF, as favored by strong lensing observations, would result in lower central dark-matter fractions for observed galaxies and can mitigate the apparent mismatch of $f_{\rm DM}$ though we note that IllustrisTNG uses a Chabrier IMF. 

To investigate whether this discrepancy between observations and simulations is the result of galaxy-formation processes, we use the publicly available Evolution and Assembly of GaLaxies and their Environment (EAGLE) hydro simulations (\citealt{s15,c15}) to perform a detailed comparison of the central dark-matter fractions, evaluated at both one-half effective radius ($R_{\rm eff}/2$) and one effective radius ($R_{\rm eff}$) from simulations and strong gravitational lensing observations of the Sloan Lens ACS Survey (SLACS; \citealt{bolton2006,koopmans2006,koopmans2009,auger2010a,auger2010b,Shu2017}). We also make use of the publicly available Illustris and IllustrisTNG simulations (\citealt{vogel2014, Wang_2019}) which have different implementations of feedback than EAGLE. 

This paper is structured as follows. In Section~\ref{simulations}, we summarize the EAGLE, Illustris and IllustrisTNG galaxy formation simulations and the relevant codes that we use in our analyses. Section~\ref{pipeline} describes the methodology used to calculate the $\fdm$. In Section~\ref{results}, we discuss the dark-matter fractions obtained from simulations and compare them with SLACS galaxies and also with other simulations. The implications of our results for galaxy formation are discussed and summarized in Section~\ref{discussions}. The values of the cosmological parameters are $\mathrm{\Omega_\Lambda}$ = 0.693, $\mathrm{\Omega_b}$ = 0.0482519, $\mathrm{\Omega_m}$ = 0.307, ${h=H_0/(100\; {\rm km\; s^{-1} \; Mpc^{-1}})}$ = 0.6777 and ${\sigma _{8}}$ = 0.8288. These are taken from the Planck satellite data release (\citealt{planck2014}), in accordance with the EAGLE, and IllustrisTNG simulations. Illustris on the other hand uses slightly different cosmological parameters, but previous analyses (\citealt{Genel2018, Pillepich2018}) have shown that these differences between Illustris and IllustrisTNG have negligible effects on the quantities explored in this work.


\section{The EAGLE, Illustris, and TNG Simulations}\label{simulations}
Here, we make use of the main Reference 100 cMpc of EAGLE, Illustris (107 cMpc) and IllustrisTNG (110 cMpc) models.
EAGLE\footnote{\url{http://icc.dur.ac.uk/Eagle/}} is a suite of hydrodynamical simulations of the formation of galaxies and other astronomical systems in a $\rm \Lambda$CDM universe (\citealt{s15,c15,McAlpine2016}). The simulations use a modified SPH (Smoothed Particle Hydrodynamics) version of {\tt GADGET 3} (\citealt{springel2005c}), with a gravitational softening length of 2.66 comoving kpc (ckpc), limited to a maximum physical scale of 0.7 proper kpc (pkpc). The initial particle masses for baryons and dark matter are $m_{\rm b} = 1.8 \times 10^6 \ \rm M_{\odot}$ and $m_{\rm dm} = 9.7 \times 10^6 \ \rm M_{\odot}$, respectively. The prescriptions for stellar and AGN feedback were calibrated to broadly reproduce the observed present-day galaxy stellar mass-function, disk sizes, and the relation between black hole and galaxy masses. The sub-grid physics includes radiative cooling (\citealt{wiersma2009}), star formation \citep{schaye2008}, stellar mass loss \citep{wiersma2009b}, thermal energy feedback from star formation \citep{vecchia2012}, black-hole accretion and AGN feedback \citep{s15,Rosas2015}. The resulting galaxies are in broad agreement with observed properties such as the star-formation rate, passive galaxy fraction, Tully-Fisher relation and colors (\citealt{s15,trayford2015}), and rotation curves (\citealt{schaller2015a}).

Illustris\footnote{\href{https://www.illustris-project.org/}{https://www.illustris-project.org/}} and  IllustrisTNG\footnote{\href{https://www.tng-project.org/}{https://www.tng-project.org/}} use the AREPO code (\citealt{springel2010}) which employs a tree-particle-mesh algorithm to solve Poisson’s equation for gravity and a second-order accurate finite-volume Godunov scheme on a moving, unstructured Voronoi-mesh for the equations of ideal magnetohydrodynamics. IllustrisTNG \citep{Nelson2019} builds upon the Illustris simulation (\citealt{vogel2014, xu2017}) and improves upon Illustris by (a) extending the mass range of the simulated galaxies and halos, and (b) adopting improved numerical and astrophysical modelling relative to Illustris \citep{Pillepich2018}.  In this work, we use the IllustrisTNG simulation box with a side length of 110.7\,cMpc having $m_{\rm b} = 1.4\times10^6\,  {\rm M}_\odot$ and $m_{\rm dm} = 7.5\times10^6\,{\rm M}_\odot$. The Illustris run also has a side length of 106.5\,cMpc and has $m_{\rm dm} = 6.26 \times 10^6\,{\rm M}_\odot$ and $m_{\rm b} = 1.26 \times 10^6\,{\rm M}_\odot$, resolving gravitational dynamics down to a physical scale of  0.710\,pkpc. IllustrisTNG reproduces the galaxy stellar mass fraction and, similar to EAGLE, has an overall agreement with observations (\citealt{Genel2018, Wang_2019}).

The simulations from EAGLE and the entire Illustris family assume a Chabrier stellar IMF (\citealt{chabrier2003}). Thus we can compare the dark-matter fractions for all three simulations on an equal footing without having to make any adjustments to their properties. 


\section{Assumptions and Data Extraction}\label{pipeline}

In this section, we discuss important assumptions pertaining to the stellar IMF and our analysis methodology. 

\subsection{The stellar initial mass function}

Currently, most observational and theoretical studies of the stellar mass in galaxies rely on assumptions about the stellar IMF which in turn determines the stellar mass per unit luminosity. 
\citet{auger2010b} found that differences in the stellar-population properties (e.g.,\ age, metallicity, extinction, star-formation history) of very massive early-type galaxies in the SLACS sample are not sufficient to account for the broad trends in their inferred total mass-to-light ratio values as a function of galaxy mass, for a non-varying (i.e.\ for a universal) IMF. Thus, \citet{auger2010b} concluded that the total mass-to-light ratio in SLACS lens galaxies increases with their total stellar mass, most likely as a result of an increasing dark-matter fraction (see also \citealt{cappellari2006, tortora2009, Thomas2011}), although the stellar IMF itself cannot be determined from these data. Hence, the analyses of SLACS lens galaxies were done for both a \cite{chabrier2003}, and a \cite{Salpeter55} IMF. In this paper we assume a Chabrier IMF for SLACS and compare with the results from EAGLE, Illustris, and IllustrisTNG where this IMF is also assumed. 
We refer the interested readers to \citet{CB2019} for EAGLE simulations with variable IMFs.

\subsection{Stellar and dark-matter masses, and effective radii}\label{sect:massreff}

We select galaxies from the simulations based on their stellar mass and produce dark matter, stellar, and gas surface mass density maps, using the SEAGLE pipeline (\citealt{mukherjee2018,mukherjee2018b}, hereafter M18 and M19 respectively), which incorporates the {\tt GLAMER} (\citealt{metcalf2014,petkova2014}) ray-tracing code and the parametric lens-modeling code {\tt LENSED} (\citealt{tessore15a,Bellagamba2017}). We infer all quantities in this paper directly from these mass maps, by-passing the lens simulations and modeling steps, which we have shown in M18 and M19 to yield very similar results. 
We set a high total galaxy stellar-mass threshold of $>10^{11}\rm M_\odot$ taken from the simulation catalogues, and excise outliers based on extreme values of their effective radius (see M18, and table~2 of M19). 
As in M18 and M19, we assume that the lens redshift is fixed at $z$=0.271 for all mock lenses, typical for median SLACS lens redshifts. We expect evolutionary effects to be small compared to the large differences and scatter that are observed between these quantities (discussed further in M18 and M19, and below). 

After extracting the particles of an individual galaxy, we project each galaxy along its three simulation coordinate axes, producing associated projected mass maps (M18). The effective radius, unlike in the observations, is not derived from the simulated galaxy brightness distribution via model fitting (e.g. \ with a S\'ersic profile), but directly inferred from the simulated stellar mass profiles as the radius enclosing one-half of the total projected stellar mass. Similarly, the stellar and total masses are derived directly from the mass maps. We use the projected half stellar mass radius as a proxy for the stellar-light effective radius (i.e.,\ $R_{\rm eff}$) for each projected mass map. We finally calculate the central projected dark matter fractions within half of the effective radius, $f_{\rm DM}(<0.5R_{\rm eff})$ for all the EAGLE early-type galaxies, following the definition in \citet{tortora2009} and \cite{auger2010b}:
\begin{equation}
\fdm \equiv 1-\frac { M_{\star}(\beta R_{\rm eff})}{ M_{\rm T}(\beta R_{\rm eff})} \; \; \; \; {\rm with}\; \beta=0.5\; {\rm or}\; 1.0,
\end{equation}
where $M_{\star}(\beta R_{\rm eff})$ is the projected stellar mass within $\beta R_{\rm eff}$ and $ M_{\rm T}(\beta R_{\rm eff})$ is the total projected mass within $\beta R_{\rm eff}$. 
Because SLACS \citep{auger2010b} does not list dark-matter fractions within $R_{\rm eff}$, we use the following equation to obtain it
\begin{equation}
    f_{\rm DM}(<R_{\rm eff})\equiv 1- \frac{\nicefrac{1}{2}\,{M_*}}{2\,{M_{\rm R_{\rm eff}/2}}},
\end{equation}
where $M_{\rm R_{\rm eff}/2}$ is the total projected mass within $R_{\rm eff}/2$, and $M_*$ is the total projected stellar mass. The total mass within $R_{\rm eff}$ is then twice that within $R_{\rm eff}/2$, assuming an isothermal model on average. The intrinsic scatter on the assumed mass-density slope is typically less than 10\% \citep{koopmans2009} and this simple extrapolation is assumed to be sufficiently accurate. 

For IllustrisTNG, we use a combination of their publicly available database\footnote{\href{https://www.tng-project.org/data/}{https://www.tng-project.org/data/}} and \citet{Wang_2019}. We obtain the effective radii and dark-matter fractions from the public database (slight differences in fig. 1 is due to manual extraction of data from Wang et al. 2019). For slopes and dark-matter fractions within $R_{\rm eff}/2$, we make use of results presented in \citet{Wang_2019}. We note that \citet{xu2017} also present the dark-matter fraction at $R_{\rm eff}/2$ and $R_{\rm eff}$ (see their Figure~11) and the $M_*$--$R_{\rm eff}$ relation from the Illustris simulations. For both Illustris and IllustrisTNG, we use the data for redshift $z=0.3$, close to the redshift of $z=0.271$ used in this work. No additional adjustments for cosmology is required because the cosmological parameters and IMF are almost identical for the three simulations used here. 


\section{Results}\label{results}
In this section, we present the results obtained from our comparisons between observed and simulated galaxies, focusing particularly on dark-matter fractions within one half and one effective radius. 

\begin{figure*}
\includegraphics[width=0.9\textwidth]{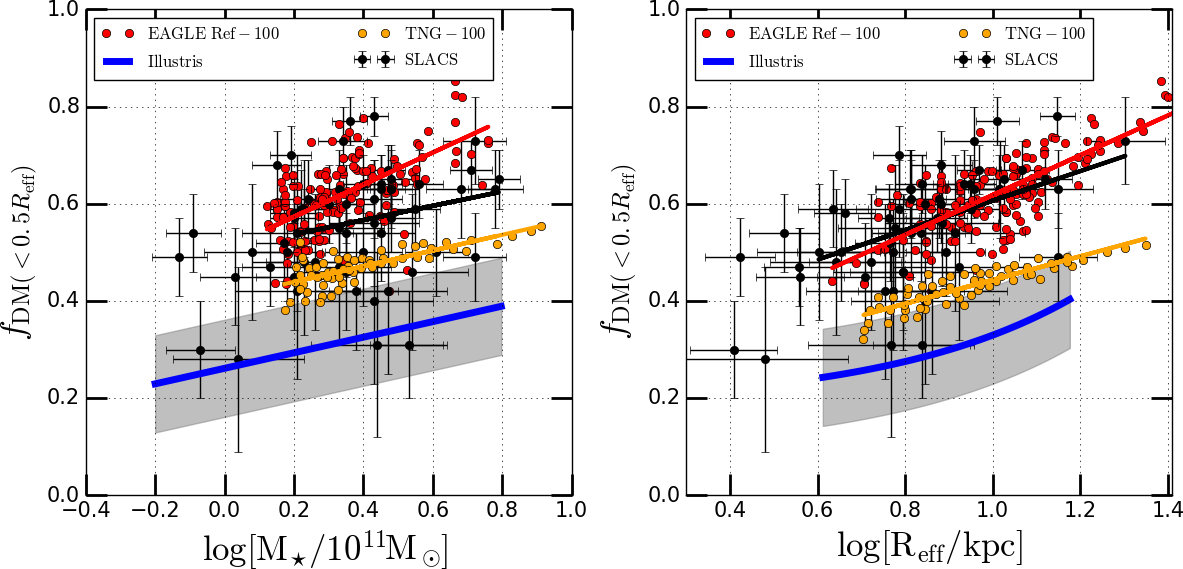}\\
\includegraphics[width=0.9\textwidth]{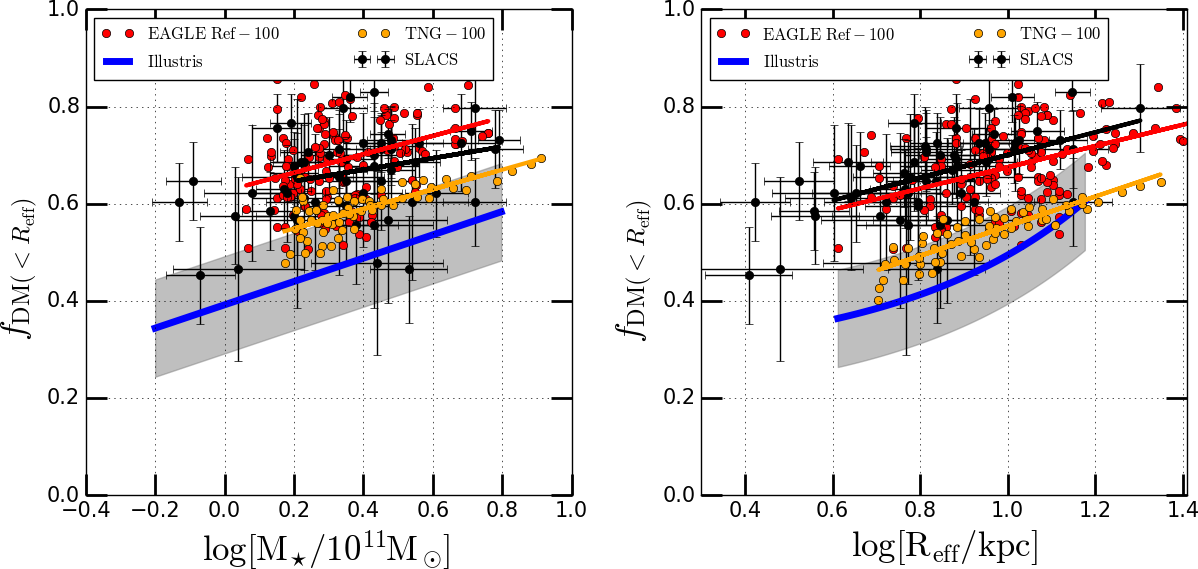}
\caption{Top row: Comparison between the EAGLE-Reference 100 cMpc simulation at $z$=0.271 (red dots), Illustris (blue line) and IllustrisTNG (orange dots) at $z$=0.3 and SLACS (black dots) for trends in $f_{\rm DM}(<0.5R_{\rm eff})$ with stellar mass, $M_{\star}$ (top left panel), and with effective radius ($ R_{\rm eff}$) (top right panel). Bottom row: Idem for $f_{\rm DM}(<R_{\rm eff})$. Both comparisons assume a constant stellar M/L ratio and a Chabrier IMF.}
\label{REF}
\end{figure*}


\begin{figure*}

\includegraphics[width=0.45\textwidth]{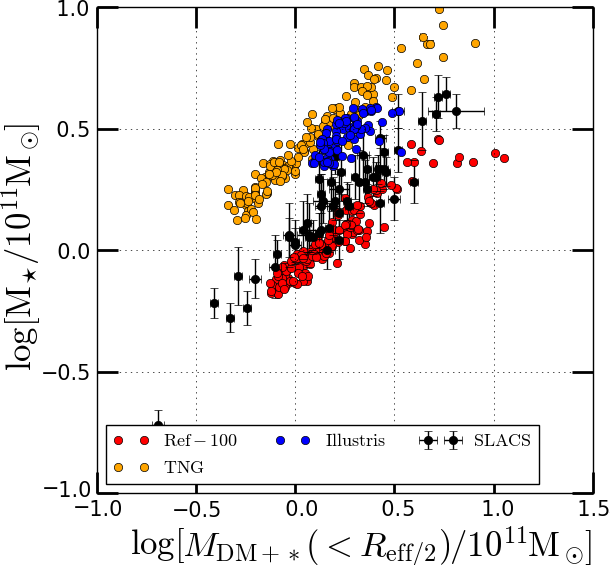}
\includegraphics[width=0.45\textwidth]{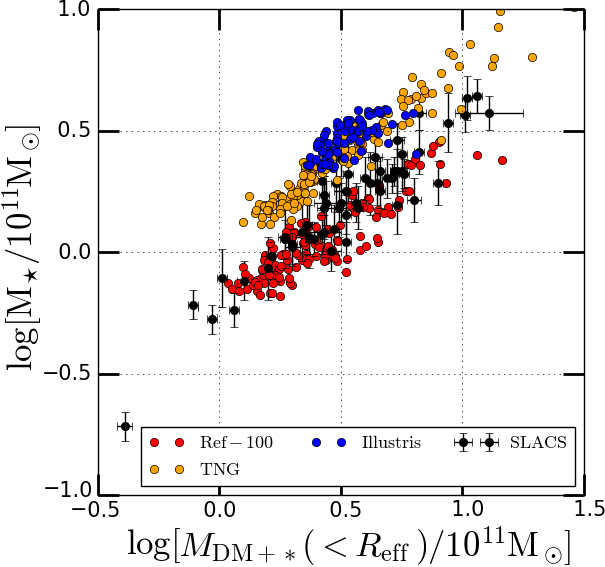}
\caption{Comparison between the EAGLE-Reference 100 cMpc simulation at $z$ = 0.271 (red dots), Illustris (blue dots), TNG (orange dots) at $z$ = 0.3, and SLACS (black dots) for trends in the combined stellar and dark matter mass ($M_{\rm DM+\star}$) at $ R_{\rm eff}/2$ with total stellar mass, $M_{\star}$ (left panel), and at $ R_{\rm eff}$ (right panel). Both comparisons assume a constant stellar M/L ratio and a Chabrier IMF.}
\label{REF_2}
\end{figure*}

\begin{figure}

\includegraphics[width=0.46\textwidth]{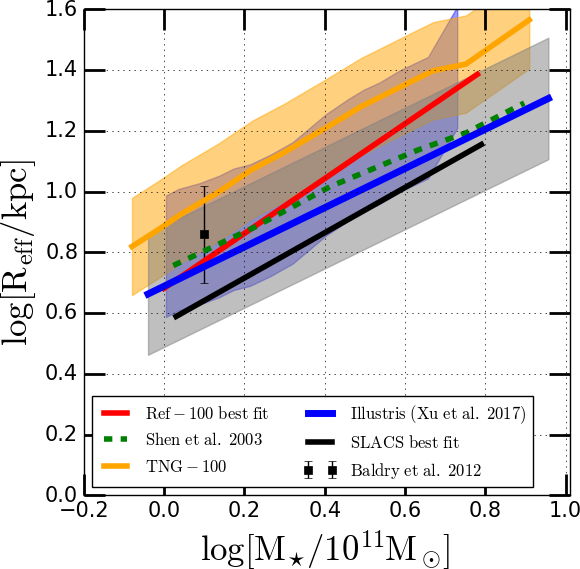}

\caption{Comparison between the EAGLE-Reference 100 cMpc simulation at $z$=0.271 (red line), Illustis (blue line), and IllustrisTNG (orange line) at $z$=0.3, and SLACS (black line) and other non lensing studies for trends in effective radius ($ R_{\rm eff}$) with stellar mass ($M_{\star}$). The shaded region is 2$\sigma$ of each distribution. The overall distribution of EAGLE galaxies are shown as blue shaded region.}
\label{REF2}
\end{figure}

 

\subsection{Dark-matter fractions inside R$_{\rm eff}/2$}\label{fdm}


The mass distribution within half of the effective radius ($R_{\rm eff}/2$) was chosen for SLACS galaxies (\citealt{koopmans2006,auger2010a,auger2010b,barnabe2011}) because it is close to the average Einstein radius and therefore leads to smaller errors when interpolating or extrapolating the mass models from the Einstein radius to this reference radius than is the case for $R_{\rm eff}$(see figure~7 in \citealt{auger2010b}). Moreover, the mass-density profile is sensitive to the scale over which it is modelled and might not represent the profile near the Einstein radius (e.g.\ \citealt{xu2017}). 
Despite this, only one recent study of hydrodynamic simulations examined $ f_{\rm DM}(<0.5R_{\rm eff})$ (\citealt{xu2017}; see figure 11 therein). 

\subsubsection{Comparing SLACS, EAGLE and Illustris(TNG)}

Figure \ref{REF} shows the trends of $f_{\rm DM}(<0.5R_{\rm eff})$ (top row) and $f_{\rm DM}(<R_{\rm eff})$ (bottom row) against two different observables ($M_{\star}$ and $R_{\rm eff}$) in the left and right column, respectively, for galaxies from EAGLE at $z_{l}$=0.271, Illustris and IllustrisTNG (both at $z_l$=0.3) and SLACS. 
We find that the $f_{\rm DM}(<0.5R_{\rm eff})$ values for SLACS and EAGLE Ref-100 are comparable, having values in the range $0.5-0.7$, whereas there is a clear mismatch with Illustris and IllustisTNG (top left panel in Figure \ref{REF}). 
%

When we plot $f_{\rm DM}(<\,0.5R_{\rm eff})$ against $R_{\rm eff}$, instead of against stellar mass, we find a tighter correlation for both EAGLE Ref-100 and IllustrisTNG (top right panel of Figure \ref{REF}). A tighter correlation  between $f_{\rm DM}(<0.5R_{\rm eff})$ and $R_{\rm eff}$, rather than with $\rm M_{\star}$, was earlier reported for SLACS (\citealt{auger2010b}). A correlation is expected in part because a larger $R_{\rm eff}$ encloses a larger portion of a galaxy's dark-matter halo (e.g.\ \citealt{tortora2012,tortora2018,xu2017,remus2017}). While EAGLE is in good agreement with the data, Illustris and IllustrisTNG predict dark-matter fractions within half of an effective radius that are appreciably lower than for SLACS, or the effective radius needs to be much larger for a given $f_{\rm DM}$ value. 
%
%
The central DM fraction values $f_{\rm DM}(<0.5R_{\rm eff})$ found in the Illustris and IllustrisTNG simulations thus differ sharply from SLACS, at fixed stellar mass (top panels in Figure \ref{REF}). 
In agreement with our results, \cite{xu2017}  calculated $f_{\rm DM}(<\,0.5R_{\rm eff})$ and $f_{\rm DM}(<\,R_{\rm eff})$ for Illustris (\citealt{vogel2014}) at $z_{l}$=0.3, and found values of $f_{\rm DM}(<\,0.5R_{\rm eff})$ lower by a factor of 2--3 than SLACS. 

However, the ratio of the projected dark-matter mass over the projected total mass can appear to be in good agreement even if both the projected stellar mass and the projected dark mass are incorrect.
Thus, in Figure \ref{REF_2} we also compare the sum of stellar and dark matter mass inside $R_{\rm eff}/2$  and $R_{\rm eff}$ to the total stellar mass. Also in this case, we find that EAGLE agrees well with SLACS, whereas Illustris and TNG galaxies do not. 

To further characterize EAGLE early-type galaxies and understand the discrepancies with other simulations, we plot their size--mass relation in the Figure \ref{REF2}. We find that the EAGLE Ref-100\,cMpc simulation yields early-type galaxies with slightly larger effective radii compared to SLACS, about 0.1--0.2 dex at similar stellar masses. In M19, we discussed possible systematics that could explain this small difference. We also compare Ref-100 with non-lensing galaxies of \citet{shen2003} and \citet{baldry2012} and find good agreement.
For IllustrisTNG, the sizes are about 0.3 dex larger than for SLACS and about 0.2 dex larger than the observations of \cite{shen2003}. The stellar mass-size relation of simulated EAGLE Ref-100 galaxies is  slightly steeper than that for Illustris galaxies but more shallow than IllustrisTNG galaxies (see also figure 2 in \citealt{Genel2018}). Thus we see that the difference in the size of a typical ETG between EAGLE, Illustris and Illustris-TNG is a few kpcs. On the other hand, the difference in dark matter fraction is almost a factor of two, which cannot be compensated with just the difference in their mass-size relation. Hence, these moderate mass-size relation differences are not the explanation for the lower dark-matter fractions in Illustris and IllustrisTNG. 



\begin{figure}
\includegraphics[width=0.45\textwidth]{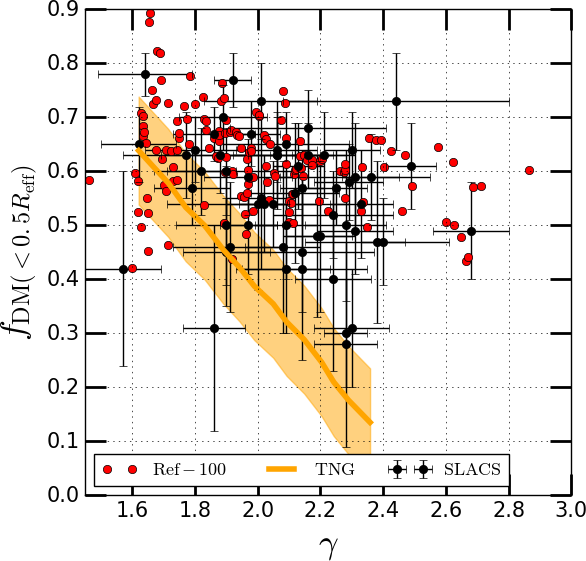}
\caption{Comparison between the EAGLE Ref-100 (red dots), IllustrisTNG (orange line) and SLACS (black dots) for trends in $f_{\rm DM}(<0.5R_{\rm eff})$ with $\gamma$. Illustris has lower DM fraction than IllustisTNG, thus its trend is not included here.}
\label{fdm_gamma}
\end{figure}

Finally, in Figure \ref{fdm_gamma}, we show the correlation between the total mass density slope ($\gamma$) and $f_{\rm DM}(<0.5R_{\rm eff})$ in EAGLE, IllustrisTNG and SLACS. Although EAGLE Ref-100 and SLACS overlap, IllustrisTNG predicts a steep decrease in dark-matter fraction with $\gamma$ that is not observed in SLACS. 




%
In a recent study, \cite{remus2017} measured $f_{\rm DM}(<R_{\rm eff})$ from the \textsc{Magneticum Pathfinder} simulations (\citealt{hirschmann2014}) and concluded that the dark-matter fractions are similar to observations of SLACS inside $R_{\rm eff}/2$. However, comparing $f_{\rm DM}$ at these two different scales can lead to $\approx 30-40\%$ differences \citep{xu2017, lovell2018}. \cite{remus2017} also did not reveal any clear correlation between stellar mass and central dark-matter fractions. 
This contrasts sharply with the  clear correlation between those quantities in both the EAGLE Ref-100 simulation and SLACS (see top panel of Figure \ref{REF}). 

\subsection{Dark-matter fractions inside R$_{\rm eff}$}


To test whether these discrepancies persist at larger radii, where finite resolution effects in the simulations are fractionally smaller, we examine the dark-matter fraction inside $R_{\rm eff}$ (see section~\ref{sect:massreff}). In the bottom row of Figure~\ref{REF}, we show the trends in $f_{\rm DM}(<R_{\rm eff})$ with stellar mass (bottom left panel) and effective radius (bottom right panel) for EAGLE Ref-100, Illustris, IllustrisTNG and SLACS. Whereas, EAGLE Ref-100 again has an overall good agreement with SLACS, 
Illustris and IllustrisTNG have considerably lower dark-matter fractions. For increasing values of $R_{\rm eff}$ and $M_*$, though, the differences decrease. Even though the difference in dark-matter fraction between IllustrisTNG and SLACS has improved from $R_{\rm eff}/2$ to $R_{\rm eff}$, it is still lower than both SLACS and EAGLE by a factor~$\approx$1.5.


\section{Discussions and Conclusions}\label{discussions}


We have investigated  the central projected dark-matter mass fraction inside a half and full effective radius of simulated galaxies with stellar masses exceeding $10^{11}$\,M$_\odot$, (a) comparing the results to the mass-selected (strong-lens) galaxies from the SLACS survey, under the assumption of a universal Chabrier stellar IMF, 
and (b) investigating trends in the dark-matter fraction with galaxy mass, size and mass-density slope for the EAGLE Ref-model, Illustris and IllustrisTNG simulations. Our main conclusions are: \\

\noindent
(i) The dark-matter fractions inside $R_{\rm eff}$/2 and $R_{\rm eff}$ found in EAGLE simulations are in good agreement with those inferred from observed SLACS galaxies. As one progresses to more massive and larger galaxies, the dark-matter fraction increases. This trend is also similar to that seen for strong-lensing observations in SLACS. EAGLE also reproduces the observed relation between dark-matter fraction and the slope of the total density profile.\\ 

\noindent
(ii) Illustris and IllustrisTNG galaxies have lower dark-matter fractions than SLACS observations and EAGLE inside both $R_{\rm eff}/2$, and $R_{\rm eff}$. We attribute this difference to differences in the subgrid feedback models, which appears to lead to an over accumulation of baryons in the inner central regions of galaxies in Illustris and IllustrisTNG.\\


The large differences in the dark-matter fractions between the Illustris and IllustrisTNG simulations and SLACS observations decrease, although do not disappear, when compared within one effective radius (see \citealt{remus2017,lovell2018}). The difference between the dark-matter fractions at $R_{\rm eff}$ and $R_{\rm eff}/2$ is too large to be attributed purely to observational and numerical resolution effects. 
%
The significant differences between EAGLE and Illustris(TNG) simulations suggest that different, or additional, mechanisms controlling star formation processes and the (re)distribution of dark matter or stars play an important role in massive early-type galaxies. 
The assumption of a universal (Chabrier) stellar IMF may be incorrect (\citealt{Dokkum2017,Smith2015}), in fact, relaxing this assumption can lead to changes in the inferred dark-matter fraction with effective radius and galaxy mass (e.g., \citep{Tortora2013} and reference therein). Moreover, low-mass stars might be more prevalent at smaller radii, mimicking dark matter concentrated in the galaxy centers in the observations (e.g., \citealt{CB2019}). However, our work suggests that we do not need to invoke a different IMF to resolve the mismatch in dark-matter fraction between simulations and observations. However, analyzing if this can affect in some way the results of the paper is not possible. In fact, if from one side many works have analyzed SLACS galaxies within a non-universal IMF scenario, simulations would require a self-consistent implementation of a variable IMF or an IMF different from the standard one. And this is not accounted for in all the analyzed simulations, preventing us from a detailed comparison.


We will investigate the impact of different types of feedback mechanisms and a non-universal IMF on the dark-matter fraction in EAGLE galaxies in a forthcoming publication.



\section*{Data Availability}
EAGLE simulations data are available at \href{http://icc.dur.ac.uk/Eagle/database.php}{http://icc.dur.ac.uk/Eagle/database.php}. Illustris and IllustrisTNG data are available at \href{https://www.illustris-project.org/}{https://www.illustris-project.org/} and  \href{https://www.tng-project.org/data/}{https://www.tng-project.org/data/} respectively. Any other requests can be made to SM/LVEK.
\section*{Acknowledgements}
We thank the anonymous referee for her/his useful comments, suggestions and helping us to improve
the paper to its current form.
SM and LVEK were supported through an NWO-VICI grant (project number 639.043.308). SM also acknowledges funding from COSMICLENS: ERC-2017-ADG, Grant agreement ID: 787886. RBM's research was supported by grant PRIN-MIUR 2017WSCC32. MS is supported by VENI grant 639.041.749. JS is supported by VICI grant 639.043.409. GV has received funding from the Marie Sklodovska-Curie grant agreement No 897124. SM thanks Fabio Bellagamba and D. D. Xu for useful discussions.



\bibliographystyle{mnras}
\bibliography{./SEAGLE-III/references.bib,ms.bib} 
\bsp	
\label{lastpage}
\end{document}